\documentclass[prl,unsortedaddress,superscriptaddress,a4paper,nofootinbib,%
nobalancelastpage,preprintnumbers,showpacs,twocolumn]{revtex4}
\usepackage{latexsym,amsbsy}
  \usepackage[dvips]{graphicx}
 \DeclareGraphicsExtensions{.eps, .jpg}
\renewcommand{\vec}[1]{\boldsymbol{#1}}
\begin{document}
\title{Stability of multiquarks in a simple string model}
\author{J.~Vijande}
\email{javier.vijande@uv.es}
\affiliation{Departamento de F\' \i sica Te\'orica and IFIC,
Universidad de Valencia - CSIC, 46100 Burjassot, Valencia, Spain}
\affiliation{Departamento de F{\'\i}sica Fundamental,
Universidad de Salamanca, 37008 Salamanca, Spain}
\author{A. Valcarce}
\email{valcarce@usal.es}
\affiliation{Departamento de F{\'\i}sica Fundamental,
Universidad de Salamanca, 37008 Salamanca, Spain}
\author{J.-M.~Richard}
\email{jean-marc.richard@lpsc.in2p3.fr}
\affiliation{Laboratoire de Physique Subatomique et Cosmologie,
 Universit\'e Joseph Fourier--INPG--IN2P3-CNRS,
53, Avenue des Martyrs, 38026  Grenoble, France}
\date{\today}

\begin{abstract}
A simple string model inspired by the strong-coupling regime of Quantum ChromoDynamics is used as a potential for studying the spectrum of multiquark systems with two quarks and two antiquarks, with a careful treatment of the four-body problem. It is found that the ground state is  stable, lying below the threshold for dissociation into two isolated mesons.
\end{abstract}

\maketitle
%
%
The question of the existence of multiquark systems is almost as old as the concept of quarks, see, e.g., \cite{Kokkedee:101899}, in particular the paper by R.H. Dalitz therein. Since the early days of hadron spectroscopy within the quark approach, many studies have been devoted to multiquark states. Of particular interest are hadrons with exotic quantum numbers that cannot be matched by any quark--antiquark ($q\bar{q}$) or three-quark ($qqq$) configuration, and among them, the states, if any, which are bound below the threshold for dissociation into two ordinary hadrons and thus are narrow and should show up clearly in the experimental spectrum.

The present contribution belongs to the class of constituent quark models: an explicit set of rules is adopted to mimic the interaction of quarks in Quantum ChromoDynamics  (QCD) and, within this  framework, the 2-body, 3-body and higher few-body  problems are solved as accurately as possible to examine whether quarks tend to split into  small ($q\bar{q}$) and  ($qqq$)  clusters or sometimes find it energetically more favorable to form multiquark clusters. After a series of estimates within the bag model (see, e.g., \cite{Aerts:1977rw}), there have been several attempts with potential models, using the powerful few-body techniques developed in atomic and nuclear physics.

Several dynamical ingredients have been identified along the years as possible sources of multiquark binding. The best known is probably chromomagnetism \cite{Jaffe:1999ze}: the spin--color operator $\vec{\sigma}_i.\vec{\sigma}_j\,\tilde{\lambda}_i.\tilde{\lambda}_j$, which is encountered in the spin-dependent part of one-gluon exchange gives rise to remarkable coherence effects, and gives in some multiquark clusters  some attraction that is larger than in its decay products. This mechanism was proposed in particular for the H-dibaryon ($uuddss$) \cite{Jaffe:1976yi}, tentatively below the $\Lambda\Lambda$ threshold, and for the 1987 version of the heavy pentaquark $(Q\bar{q}\bar{q}\bar{q}\bar{q})$ \cite{Gignoux:1987cn}. The chromomagnetic scenario has, however,  difficulties: the first optimistic predictions carried out in the limit of exact flavor SU(3) symmetry, and using short-range correlation coefficients borrowed from ordinary hadrons, do not survive a more careful dynamical treatment \cite{Rosner:1985yh}.

Another binding mechanism is based on the flavor-independence of the confining interaction. In a given static potential $V(\vec{r}_1, \ldots)$, the asymmetric mass configurations $(QQ\bar{q}\bar{q})$ tend to be lower than the threshold $2(Q\overline{q})$ if the mass ratio is large enough \cite{Ader:1981db}. This is the same favorable breaking of symmetry which makes the hydrogen molecule much more stable than the positronium molecule, in the case where the potential is taken as the Coulomb interaction (see, e.g., \cite{ARV} for references).

Now, the determination of the critical mass ratio $M/m$ at which $(QQ\bar{q}\bar{q})$ becomes stable, and the existence of other multiquark systems  depend crucially on  questionable assumptions on the multiquark potential. However successful is a potential $v(r)$ for the spectrum of quarkonium, its extrapolation to baryons and multiquarks remains, indeed, somewhat risky.

\begin{figure}[!htb]
\centerline{\includegraphics[width=.45\textwidth]{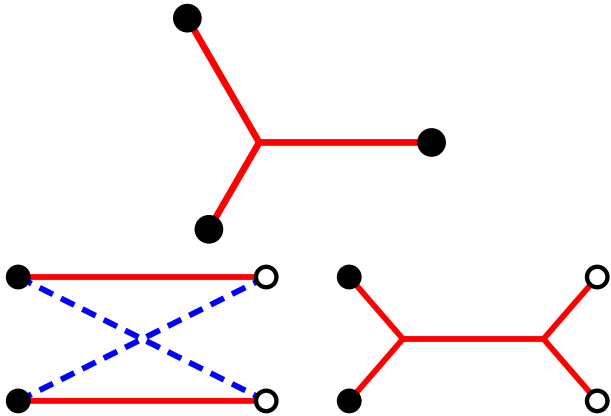}}
\caption{\label{inter:fig:st} String model for three-quark confinement (top), and for four quarks (bottom): flip-flop (left) and  ``butterfly" (right), an alternative configuration that is favored when the quarks (full disks) are well separated from the antiquarks (open circles).}
\end{figure}

There are interesting attempts \cite{Stanley:1980fe} to describe mesons and baryons simultaneously with the potential energy of the latter systems  taken as 
\begin{equation}\label{intro:eq:half-rule}
V(\vec{r}_1,\vec{r}_2,\vec{r}_3)={1\over2}\left[v(r_{12})+v(r_{23})+v(r_{31})\right]~,
\end{equation}
where $r_{ij}$ is the relative distance between particles $i$ and $j$. It is tempting to extrapolate this potential as 
\begin{equation}\label{intro:eq:l-l-rule}
V(\vec{r}_1,\ldots)=-{3\over 16}\sum_{i<j} \tilde{\lambda}_i.\tilde{\lambda}_j v(r_{ij})~,
\end{equation}
to higher multiquark systems, and benefit from the few-body techniques with pairwise potentials. This was the basis of most multiquark calculations, so far.
However, the success of the ansatz (\ref{intro:eq:half-rule}) is probably accidental, since there are many indications that if the quark-antiquark confinement is linear, $v(r)=\lambda r$, the true confining interaction for three quarks in a baryon is more likely the so-called $Y$-shape potential \cite{Artru:1974zn,Takahashi:2000te}
\begin{equation}\label{intro:eq:Y-pot}
Y(\vec{r}_1,\vec{r}_2,\vec{r}_3)= \lambda \min_k (r_{k1}+r_{k2}+r_{k3})~,
\end{equation}
where the sum of distances from a junction $k$ to the three quarks is minimized, as in the well-known problem of Fermat and Torricelli, also reported to as the Steiner problem \cite{Steiner}.  It is schematically represented in Fig.~\ref{inter:fig:st}.  From this point of view, the success of the empirical model (\ref{intro:eq:half-rule}) comes from the perimeter in a triangle being nearly equal to twice the minimal sum of distances $Y$.

The analog of (\ref{intro:eq:Y-pot}) for four-quark systems is rather complicated. Some diagrams generalizing the $Y$-shape string have been drawn in the time of baryonium (see, e.g.,  \cite{Chan:1978nk}), but they were not followed by any thorough four--body calculation. 
The dynamics of systems made of two quarks and two antiquarks has been discussed by
by several authors, for instance \cite{Dosch:1982ep,Lenz:1985jk,Carlson:1991zt}, where we found guidance. The model we use for an  exploratory study  is restricted to the sole confinement.  The short-range Coulomb-like interaction, as well as the spin-dependent terms are neglected altogether. We also take the non-relativistic limit.
The interaction which is adopted here for a careful estimate of the four-body energy 
is a combination of the string limit used for mesons and for baryons,
and will be referred to as the ``string potential".  It is schematically pictured in Fig.~\ref{inter:fig:st}
and it reads,
\begin{equation}\label{inter:eq:str}
V_{s}=\min(V_{f},V_{b})~.\,\,
\end{equation}
$V_f$ stands for the so-called ``flip-flop" model
\begin{equation}\label{inter:eq:ff}
V_{f}=\lambda \min( r_{13}+r_{24},r_{23}+r_{14})~,
\end{equation}
where each of the quarks 1 or 2 links to either antiquarks 3 or 4, to minimize the sum of the two terms.
It is understood here that the gluon field readjusts immediately to its minimal configuration. This is a kind of ideal Born--Oppenheimer limit.
$V_b$ is the butterfly-like configuration, where two quarks form a color $\bar{3}$ diquark, two antiquarks a color $3$ antidiquark, which are linked by a flux tube, 
\begin{equation}\label{inter:eq:bu}
V_{b}=\lambda \min_{k,\ell} (r_{1k}+r_{2k}+r_{k\ell}+r_{\ell 3}+r_{\ell 4})~.
\end{equation}

For the $Y$-shape potential of baryons, the junction achieving the minimal energy is either one of the quarks, if the triangle linking the three quarks is flat (an angle larger than $120^\circ$) or the point from which all sides are seen under $120^\circ$. This gives an explicit expression in terms of the interquark distances \cite{Takahashi:2000te}.
In the case of the butterfly potential, some rigorous geometrical properties remain, but one cannot avoid
 some numerical minimization over some of the junction coordinates, and this slows down the four-body variational computation.

Our numerical results have been obtained using correlated Gaussians, namely, if
\begin{equation}\label{fbp:eq:jaco}
\vec{x_1}=\vec{r}_2-\vec{r}_1~,\quad
\vec{x_2}=\vec{r}_4-\vec{r}_3~,\quad
\vec{x_3}={1\over2}(\vec{r}_3+\vec{r}_4-\vec{r}_1-\vec{r}_2)~,
\end{equation}
are the Jacobi variables, suitably generalized in the case of unequal masses, 
the trial wave function is sought  as
\begin{equation}\label{fbp:eq:gauss}
\Psi(\vec{x_1},\vec{x_2},\vec{x_3})=\exp\left[-\sum_{i\geq j=1}^3 a_{ij}\,\vec{x}_i\cdot\vec{x}_j\right]+\cdots~,
\end{equation}
where the ellipses are meant for terms deduced from the first Gaussian to restore the proper symmetry of the whole wave function.
This method is widely used in nuclear physics and \textsl{ab-initio} quantum chemistry \cite{Suzuki:1998bn}. The range parameters $a_{ij}$ are optimized numerically.

For cross-check, we also use an exponential function borrowed from the famous paper in which the stability of the positronium molecule was demonstrated \cite{PhysRev.71.493}, namely
\begin{equation}\label{fbp:eq:exp}
\Phi=\exp\left[-a (r_{13}+r_{24})-b(r_{14}+r_{23})\right]+ \cdots~,
\end{equation}
but the calculation has been restricted to the case  of the flip-flop interaction.

Note that since the potential is proportional to the distance, the virial theorem states that the kinetic, $\langle K \rangle $, and potential, $\langle V\rangle $, contributions to the energy $E=\langle K \rangle+\langle V\rangle$ are in ratio $\langle V\rangle=2 \langle K \rangle$.  This also holds for the variational energy if the space of trial functions is globally invariant under rescaling, see, e.g., \cite{ARV}. Hence instead of minimizing  $\langle K \rangle+\langle V\rangle$ with, say, $n$ parameters, it is sufficient to minimize $(4 \langle K \rangle \langle V\rangle^2/27)^{1/3}$ with $n-1$ parameters.

By scaling,  the string constant can be set to $\lambda=1$ without loss of generality, and one of the masses also taken as the unit of mass, $m=1$. 
In these units, the ground-state energy of a meson with both quark and antiquark of mass $m=1$ is 
$E_2(1,1)=2.338$ (the opposite of the first zero of the Airy function), and for a meson of masses $m_1$ and $m_2$, it is 
\begin{equation}\label{resu:eq:thr}
E_2(m_1,m_2)=E_2(1,1) \left(2 m_1 m_2\over m_1+m_2\right)^{-1/3}~.
\end{equation}
We first consider the case of equal masses, in the simple flip-flop model with exponential functions.
The meson energy, if calculated variationally from $\phi(r)=\exp(- \alpha r)$, $\alpha$ being adjusted, is
$E_2= 2.476$. This is not very good, since the cusp in this wave function is absent from the exact wave function. With a combination of two such exponentials, the meson energy is improved to $E'_2=2.353$.  Now with the wave function (\ref{fbp:eq:exp}), and the flip-flop potential, the minimal energy for the $(qq\bar{q}\bar{q})$ system is found at $E_4=4.872$. The observation that $E_4<2 E_2$, corresponding to stability within the simplest approximation in each sector,  is an indication that the flip-flop potential tends to bind the system.

\begin{table}[t]
\caption{\label{resu:tab:Mm} Four--quark variational energy $E_4$ of $(QQ\overline{q}\overline{q})$ for the different confinement models described in 
Eq. (\ref{inter:eq:str}),  compared to its threshold, $T_4=2E_2(1,M)$, and variational energy $E'_4$ of $(Q\overline{Q}q\overline{q})$ with the flip-flop model $V_f$, compared to its threshold $T'_4=E_2(M,M)+E_2(1,1)$, as a function of the mass ratio.}
\begin{ruledtabular}
\begin{tabular}{ccccccc}
$M/m$	& \multicolumn{3}{c}{$E_4$}	& $T_4$ & $E'_4$ & $T'_4$\\ 
	& $V_f$	& $V_b$	& $V_s$	&  & $V_f$  &\\ 
\hline
1	& 4.644	& 5.886	& 4.639	& 4.676  &  4.644 & 4.676 \\
2	& 4.211	& 5.300	& 4.206	& 4.248  & 4.313  & 4.194\\
3	& 4.037	& 5.031	& 4.032	& 4.086  & 4.193  & 3.959 \\
4	& 3.941	& 4.868	& 3.936	& 3.998  & 4.117  & 3.811\\
5	& 3.880	& 4.754	& 3.873	& 3.942  & 4.060  & 3.705\\
\end{tabular}
\end{ruledtabular}
\end{table}

\begin{figure}[!thb]
\begin{center}\includegraphics[width=.35\textwidth]{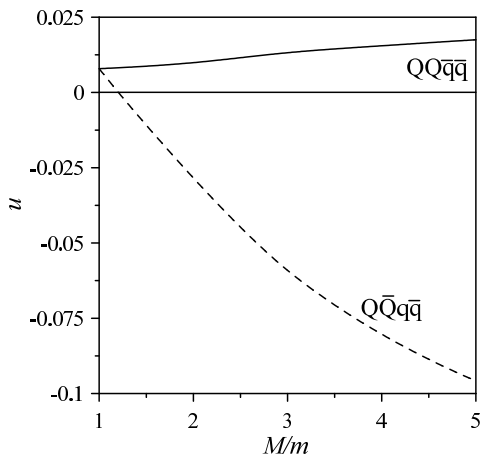}\end{center}
\caption{\label{fig3}Dimensionless excess of binding with respect to the threshold, $u$, as a function of the mass ratio $M/m$.}
\end{figure}

If this computation is now carried out with the Gaussian wave function (\ref{fbp:eq:gauss}), one obtains
a variational energy $E_4=4.644$, which is below the threshold for spontaneous dissociation. The string model slightly lowers this energy, to $E_4=4.639$.

If the calculation is done with the trial wave function (\ref{fbp:eq:gauss}), but with the restriction $i=j$ which implies 
that only $x_i^2$ terms are allowed, the minimum of the energy is found appreciably larger, $E_4$=4.797 for $M/m=1$ and $E_4$=4.342 
for $M/m=2$. Comparing these results with Table \ref{resu:tab:Mm}, the four--quark states would be above the dissociation threshold. This approximation, which consists of neglecting the relative angles between the different Jacobi coordinates and therefore internal 
relative orbital angular momentum, was used in Ref.\ \cite{Carlson:1991zt} with a similar ``string potential'', leading the authors to conclude
that no bound exotic states exist (even for the sole confinement potential, see Table I of Ref.\ \cite{Carlson:1991zt}).

We now introduce some symmetry breaking in the kinetic energy, and consider the configurations $(QQ\bar{q}\bar{q})$ and $(Q\overline{Q}q\bar{q})$ with two different masses, $M$ and $m$. The results are shown in Table \ref{resu:tab:Mm}. Clearly, as $M/m$ increases, a deeper binding is obtained for the flavor-exotic  $(QQ\bar{q}\bar{q})$ system. For the hidden-flavor $(Q\overline{Q}q\bar{q})$, however, the stability deteriorates, and with our variational approximation, for $M/m\gtrsim 1.2$, the system becomes unbound with respect to its lowest threshold $(Q\overline{Q})+(q\bar{q})$\footnote{An energy above the threshold simply means that the system is unbound within our variational approximation, suggesting that the minimum of the Hamiltonian is at the two-meson threshold. It would be more difficult to find an approximate mass for a possible meson--meson resonance.}. The amount of binding, independent of any scale factor, is well measured by the dimensionless coefficient defined by
\begin{equation}\label{resu:eq:udef}
E_4=(1-u)T_4~,
\end{equation}
linking two-body and four-body energies. A plot of $u$ is given in Fig.~\ref{fig3}, as a function of the mass ratio $M/m$.

Our main conclusions and comments are in order:
\begin{itemize}\itemsep -3pt
\item With a string model including four-body forces, inspired by the strong coupling regime of QCD, the ground-state energy of the system made of two quarks and two antiquarks of equal masses is found below the dissociation threshold.
\item For the flavor exotic $(QQ\bar{q}\bar{q})$, binding becomes better when the mass ratio increases.
\item
For the cryptoexotic $(Q\overline{Q}q\bar{q})$, the effect of symmetric breaking is opposite. In atomic physics, while $(p,p,e^-,e^-)$ is more stable than the positronium molecule, the configuration $(M^+,M^-,m^+,m^-)$ becomes unstable (besides internal annihilation) for $M/m>2.2$ \cite{ARV}.
\item
For $(1,2,3,4)=(QQ\bar{q}\bar{q})$ configurations, we considered quarks (or antiquarks) of equal mass, for the ease of computation. We neglected the effect of statistics, i.e., our results directly apply to quarks (or antiquarks) having different flavor and about the same mass.  Then the interquark interaction is really an effective potential with the gluon degrees of freedom integrated out, the analog of the nucleus--nucleus effective interaction for diatomic molecules in atomic physics.
For genuinely identical quarks, another approach is possible, where each component of the interaction is associated to a specific color wave function, for instance $(1,2)_{\bar{3} }(3,4)_{3}$ for the ``butterfly potential, $(1,3)_1(2,4)_1$ or $(1,4)_1(2,3)_1$ for each component of the ``flip-flop'', in an obvious notation. A formalism has been developed in Ref.~\cite{Lenz:1985jk}, but it was associated to a quadratic interaction, and hence the results are not directly comparable to ours.
\item
The stability of four-quark states is demonstrated using a rather simple wave function. However, the dependence upon the angle between the Jacobi variables is crucial. Its neglect explains why stability was missed in earlier investigations.
\item
It is delicate to compute the connected-string contribution (butterfly) to the potential, but this is not rewarding, as the dynamics of binding is dominated by the simple flip-flop term.
\item
It would be interesting to analyze the results of lattice QCD in terms of the strength parameters associated to our string potential, and also in terms of departures from this simple ansatz. For a discussion on the multiquark interaction on lattice, see, e.g.,~\cite{Green:1993ag}.
\end{itemize}

In brief, the question of saturation raised in the early days of the quark model looks even more open today.  Chromoelectric models based on simple pairwise forces (\ref{intro:eq:l-l-rule}) do not bind tetraquarks, except in the limit of high mass ratios for $(QQ\bar{q}\bar{q})$.
Our result indicates that  with a more plausible scenario for the spin-independent potential, the starting point is \emph{stability}.

The present study is focused on the role of confining forces alone, to demonstrate that a string model of confinement leads to results which differ qualitatively from these obtained from additive pairwise potentials.
It remains to examine whether the necessary refinements will spoil or improve this binding. Among them, let us mention: relativistic effects, spin-dependent terms, Fermi statistics for identical quarks, long-range Yukawa forces between clusters (by itself this mechanism might produce binding, as shown in the ``molecular'' models  of the X(3872) \cite{Swanson:2006st}). Although the many-body confinement forces  discussed in this paper
play a role for any four-quark system, their
contribution to generate binding should be more
evident for quarks of the second generation.
For light quarks or antiquarks, chromomagnetic
effects or their analogs in chiral dynamics are
important and presumably  dominate the issue of
stability; if the threshold includes a pion, it
is obviously difficult to imagine a four-quark
state below that threshold. For very heavy
quarks, the Coulomb term dominates the
spin-independent interaction: the problem of
stability belongs then to well-studied class of
models with additive pairwise forces \cite{Ader:1981db}.
It is hoped that the encouraging results obtained
with the string model for confinement  will stimulate
intensive investigations of multiquark systems in
more refined constituent quark models with
phenomenological applications to hadron
spectroscopy. It is our intent to participate to
this enterprise.

\end{document}